%% file: paper.tex
\documentclass[aps,prl,twocolumn,showpacs,superscriptaddress,letterpaper,asmmath,amssymb]{revtex4}  

\usepackage{graphicx}
\usepackage{dcolumn}
\usepackage{bm}

\RequirePackage{xspace}
\usepackage{relsize}
\def\CP                {\ensuremath{C\!P}\xspace}

\def\Wpm    {\ensuremath{W^\pm}\xspace}
\def\Wmp    {\ensuremath{W^\mp}\xspace}

\def\t     {\ensuremath{t}\xspace}
\def\tbar  {\ensuremath{\overline t}\xspace}
\def\piz   {\ensuremath{\pi^0}\xspace}

\def\pim   {\ensuremath{\pi^-}\xspace}
\def\Kbar  {\kern 0.2em\overline{\kern -0.2em K}{}\xspace}

\def\Kp    {\ensuremath{K^+}\xspace}

\def\KS    {\ensuremath{K^0_{\scriptscriptstyle S}}\xspace} 
\def\B       {\ensuremath{B}\xspace}
\def\Bbar    {\kern 0.18em\overline{\kern -0.18em B}{}\xspace}

\def\Bz      {\ensuremath{B^0}\xspace}

\def\Bu      {\ensuremath{B^+}\xspace}

\def\Bp      {\ensuremath{\Bu}\xspace}

\def\Bs      {\ensuremath{B_s}\xspace}

\def\jpsi     {\ensuremath{{J\mskip -3mu/\mskip -2mu\psi\mskip 2mu}}\xspace}

\def\tprime  {\ensuremath{t^{\prime}}\xspace}
\def\bprime  {\ensuremath{b^{\prime}}\xspace}

\def\TFT     {\ensuremath{T_{5/3}}\xspace}
\def\missET {{\not\!\! E_T}}
\def\Qbar    {\kern 0.08em\overline{\kern -0.08em Q}{}\xspace}

\def\invfb   {\ensuremath{\mbox{\,fb}^{-1}}\xspace}
\newcommand{\mev}{\ensuremath{\mathrm{\,Me\kern -0.1em V}}\xspace}
\newcommand{\mevc}{\ensuremath{{\mathrm{\,Me\kern -0.1em V\!/}c}}\xspace}
\newcommand{\mevcc}{\ensuremath{{\mathrm{\,Me\kern -0.1em V\!/}c^2}}\xspace}
\newcommand{\gev}{\ensuremath{\mathrm{\,Ge\kern -0.1em V}}\xspace}
\newcommand{\gevc}{\ensuremath{{\mathrm{\,Ge\kern -0.1em V\!/}c}}\xspace}
\newcommand{\gevcnospace}{\ensuremath{{\mathrm{\,Ge\kern -0.1em V\!/}c}}}
\newcommand{\gevcc}{\ensuremath{{\mathrm{\,Ge\kern -0.1em V\!/}c^2}}\xspace}

\def\missET {{\not\!\! E_T}}

\begin{document}
\bibliographystyle{apsrev}

\title{{Search for Fermion-Pair Decays $Q\Qbar \rightarrow
  (\t\Wmp)(\tbar\Wpm)$ in Same-Charge Dilepton Events}}

\input October2009_Authors.tex  

\begin{abstract}
\input abstract
\end{abstract}



\maketitle

\date{\today}

The standard model (SM) of particle physics accomodates three generations
of fundamental quarks and leptons, but does not prohibit a
fourth. Recent measurements of charge-parity (\CP) nonconservation in
\B-meson decays~\cite{cpvnote} have two features in common: they are more than two standard deviations
from SM expectations, and they are sensitive to contributions~\cite{predictref}
from a fourth-generation up-type quark, \tprime.  This pattern of
measurements~\cite{phisref,Barberio:2008fa,acpref,stbref}, if genuine,
warrants a search for another generation of quarks or a multiplet of
quark-like objects. Another quark generation
(Cf.\ \cite{Frampton:1999xi}) could
provide sources of particle-antiparticle
asymmetries large enough to account for the baryon asymmetry of the
universe~\cite{Hou:2009BAU}, and accomodate a heavier Higgs boson
(the source of electroweak symmetry breaking and mass generation)
than a three-generation model~\cite{Kribs:2007nz}.

This Letter reports a search for heavy particles $Q$ decaying to a top
quark and a $W$ boson at a mass scale relevant to both the $B$-meson
anomalies and the Higgs mechanism. We search for pair-production of $Q\Qbar$ via strong interactions,
where $Q$ is either a fourth-generation down-type quark \bprime or
a quark-like (non-hadronic) composite fermion $B$ or \TFT~\cite{Contino:2008hi}.
The $B$ and \TFT\ (with 5/3 electron charge) that we consider might arise
from symmetries, consistent with precise electroweak
measurements~\cite{Sikivie:1980hm,Agashe:2006at}. If \TFT\ exists, the existence of $B$ is implied, doubling the expected
event rate.  Many additional models of new phenomena that provide a Higgs
mechanism also predict particles with large couplings to the
third-generation $t$ quark and similar decay modes.  For instance, models of
warped extra dimensions, equivalent to models of strongly interacting
composite particles, predict fermion excitations with the quantum
numbers of quarks. A summary is given in \cite{Contino:2006nn}.

In each case, $Q \to t W^{\mp}, t \to b W^+$~\cite{conjugates}.
 We investigate the case in which two
same-charge $W$ bosons decay leptonically (including $\tau$ decays to
$e$ or $\mu$).  This is the first search for quark-like particles in
this mode~\cite{previous}, achieving the most sensitive direct limits on short-lived
fourth-generation particles.  (We do not consider long-lived
particles, since displaced
vertices require different analysis methods.)

We assume that $Q$ decays exclusively to $t W^{\mp}$.
This is expected for \B\ and \TFT,  and for \bprime
it is expected~\cite{Hung:2007ak} under the assumptions that
(a) coupling to light quarks is insignificant,
(b) $m_{\bprime} >$ $m_{top}+m_{W}$=255\gevcc, and 
(c) $|m_{\tprime} - m_{\bprime}| < m_W$.
These assumptions are justified by experimental constraints.
A search for $Q \to W$+jet~\cite{Lister:2008is} found
$m_{\tprime} >$ 311\gevcc, implying
a similar limit on $m_{\bprime}$
if the \bprime\ decay to this channel is significant.
Combining this limit with results of a search for
$Q \to Z$+jets~\cite{Aaltonen:2007je} 
and an analysis of branching fractions for
\bprime~\cite{Hung:2007ak}, we infer $m_{\bprime} >$ 255\gevcc.
A fourth generation is
most consistent with precise electroweak measurements
when the mass splitting $\Delta m$ between \bprime\ and \tprime
is less than the $W$-boson mass but non-zero; ref. ~\cite{Kribs:2007nz} gives
$\Delta m \approx$ 50\gevcc, based on  $|m_{\tprime} - m_{\bprime}| < m_W$.

We use a data sample corresponding to an integrated luminosity of 2.7\invfb
collected with the CDF II detector~\cite{cdf} at the Tevatron $p\overline{p}$
collider at Fermilab.
The data acquisition system is triggered by $e$ or $\mu$ candidates
with $p_T >$ 18\gevc~\cite{coordinates}.
We require the  $\ell^{\pm}\ell^{\pm}bj\missET$
signature, following \cite{Aaltonen:2008hx}:
two same-charge reconstructed leptons ($e$ or $\mu$) with
pseudorapidity magnitude 
$|\eta| <$ 1.1 and $p_T >$ 20\gevc, where at least one lepton is
isolated~\cite{isodef}; at least two jets with $E_T >$ 15\gev\ and $|\eta| <$ 2.4;
at least one of the jets with evidence of a long-lived particle
($b$-tag) using the tight {\sc secvtx} algorithm~\cite{secvtx}; and
missing transverse energy $\missET >$ 20\gev~\cite{metdef}.

The dominant background comes from events in which one of the leptons
is a misidentified light-flavor jet or a lepton from the decay of a
bottom or charmed hadron in a heavy flavor jet, largely from $W$ production in association with light or heavy flavor
jets or from $t\bar{t}$ production with
semi-leptonic decays.  This background is described using a
lepton misidentification model from inclusive jet data~\cite{ttbar}
applied to $W+$jet events.  In same-charge dilepton control regions
without a $b$-tag requirement, this model decribes well the kinematics
of observed events with large missing transverse energy.  Nevertheless, the
requirement of a $b$-tag in the final selection introduces uncertainty
regarding the misidentification model, leading to
a final 100\% systematic uncertainty, as described in ~\cite{Aaltonen:2008hx}.

Other backgrounds include processes that
produce electron-positron pairs.  These may be reconstructed with the
same charge due to asymmetric $\gamma$ conversions in the process
$e^-_{\rm hard}\rightarrow e^-_{\rm soft}\gamma\rightarrow
e^+_{\rm hard} e^-_{\rm soft} e^-_{\rm soft}$, where hard and soft
refer to large and small transverse momentum, respectively.
The major contributions from this mechanism are from events with a $Z$
or virtual $\gamma$ in association with jets ($Z/\gamma^*+$jets) and $t\bar{t}$ production with fully leptonic decays.

\begin{table}[b]
\vspace{-3ex}
\caption{Expected background contributions to the $ee$, $e\mu$, and
  $\mu\mu$ channels in 2.7 fb$^{-1}$ from (a) $Z$ and diboson,
  (b) $ t\bar{t}\rightarrow \ell^+\nu b\ell^-\nu \bar{b}$, and
  (c) misidentified lepton.}
\begin{ruledtabular}
\begin{tabular}{lrclrclrclrcl}
Source  & \multicolumn{3}{c}{$ee$}  & \multicolumn{3}{c}{$\mu\mu$}  & \multicolumn{3}{c}{$e\mu$} & \multicolumn{3}{c}{Total $\ell\ell$} \\
\hline
(a)   &0.01&\hspace{-0.5em}$\pm$&\hspace{-0.5em}0.01& 0 & & & 0.02&\hspace{-0.5em}$\pm$&\hspace{-0.5em}0.02 & 0.03&\hspace{-0.5em}$\pm$&\hspace{-0.5em}0.03 \\
(b)   & 0.06&\hspace{-0.5em}$\pm$&\hspace{-0.5em}0.04  & 0& & &0.09 &\hspace{-0.5em}$\pm$&\hspace{-0.5em}0.03  & 0.15&\hspace{-0.5em}$\pm$&\hspace{-0.5em}0.05 \\
(c)    & 0.6&\hspace{-0.5em}$\pm$&\hspace{-0.5em}0.6  & 0.3&\hspace{-0.5em}$\pm$&\hspace{-0.5em}0.3  & 0.5&\hspace{-0.5em}$\pm$&\hspace{-0.5em}0.5  & 1.4& \hspace{-0.5em}$\pm$&\hspace{-0.5em}1.4 \\
\hline
Total                   & 0.7&\hspace{-0.5em}$\pm$&\hspace{-0.5em}0.6  & 0.3&\hspace{-0.5em}$\pm$&\hspace{-0.5em}0.3  & 0.6 &\hspace{-0.5em}$\pm$&\hspace{-0.5em}0.5 & 1.6& \hspace{-0.5em}$\pm$&\hspace{-0.5em}1.4\\
Data            & 0&&     & 1&&     & 1&&     & 2&&\\
\end{tabular}
\label{table:back}
\end{ruledtabular}
\end{table}

\begin{table}[b]
\vspace{-3ex}
\caption{Theoretical cross-sections ($\sigma_{\rm NLO}$ in fb~\cite{nloa,nlob}),
  expected yield ($N$), median expected 95\% C.L
  limit ($\sigma_{\rm exp'd}$ in fb), and observed  95\% C.L  limit ($\sigma_{\rm
    obs}$ in fb)  for \bprime (or $B$) and (\TFT\ $+B$) signals at
  varying masses.}
\begin{ruledtabular}
\begin{tabular}{lcrrrrrrrrrrrrrr}
\multicolumn{2}{c}{Mass [GeV/$c^2$]} & 300 & 310 & 320 & 330 & 340 & 350 & 375 & 400  \\
\hline
&$\sigma_{\rm NLO}$  &227    &176    &137    &106    &83     &64 & 34 & 18 \\
$\bprime or B$&$N$ &  13.4 & 9.6 & 7.5 & 5.9 & 4.6 & 3.5 & 1.9 & 1.0 \\
 & $\sigma_{\rm exp'd}$  &67 & 63 & 63 & 62 & 63 & 63 & 63 & 57      &\\
   & $\sigma_{\rm obs}$   &67 & 96 & 83 & 94 & 85 & 83 & 78 & 67   &\\ \hline
&$\sigma_{\rm NLO}$  & 454 & 352 & 274 & 212 & 166 & 128 & 68 & 36 \\
\raisebox{0ex}[0pt][0pt]{\parbox[t]{0.2cm}{$\TFT$\\\hspace*{0.5em}$+B$}}&$N$ &  27.0 & 19.5 & 15.3 & 11.9 & 9.4 & 7.1 & 3.6 & 2.1\\
 &$\sigma_{\rm exp'd}$   &86 & 89 & 69 & 62 & 59 & 65 & 66 & 60	\\

& $\sigma_{\rm obs}$  & 86 & 89 & 69 & 98 & 91 & 83 & 83 & 79 & 	\\ 

\end{tabular}
\label{table:bprres}
\end{ruledtabular}
\end{table}

\begin{figure}[t]
\includegraphics[width=0.8\linewidth]{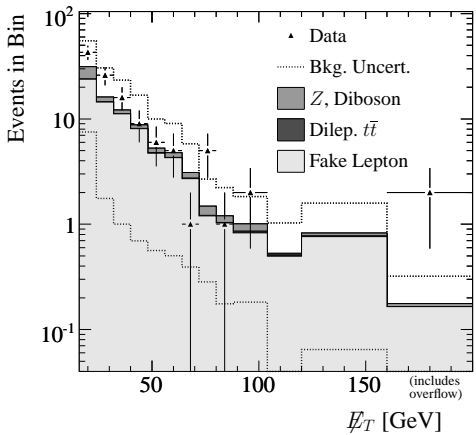}
\caption{Missing transverse energy in events with same-charge
  leptons in 2.7\invfb\ . The right outer-most bin includes overflow
  events with $\missET > 160$ GeV.
  \vspace{-3ex}}
\label{fig:metplots}
\end{figure}

Estimates of the backgrounds from $Z/\gamma^*$+jets processes
are made with the {\sc alpgen}~\cite{alpgen} v2.10 simulation code
interfaced with {\sc pythia} 6.325~\cite{pythia} in the MLM
scheme~\cite{alpgen} for the hadronization and fragmentation
and normalized to data in opposite-charge events in the $Z$ mass region.    The detector response for both $Z$+jets and $t\overline{t}$ processes is
evaluated using the CDF simulation program {\sc cdfsim}~\cite{cdfsim}, where, to avoid double-counting, the same-charge leptons are required
to originate from the $W$ or $Z$ decays rather than from misidentified
jets.

To validate the modeling of the rate of hard bremsstrahlung from electrons, we compare our prediction for
the contribution of $Z\rightarrow e^+e^-$ to the observed sample of same-charge electrons
or positrons without a $b$-tag or missing transverse energy
requirement. The shape of the dilepton invariant mass spectrum and
yield in the $Z$ mass region ($M_{ll} = [M_Z - 20, M_Z+20]$) agrees well with the prediction. In addition, $\mu\mu$ and $e\mu$
events have negligible contributions from hard bremsstrahlung, as
predicted. Figure~\ref{fig:metplots} shows that the missing
transverse energy in inclusive same-charge dilepton events is well described.

\begin{figure}[t]
\includegraphics[width=0.8\linewidth]{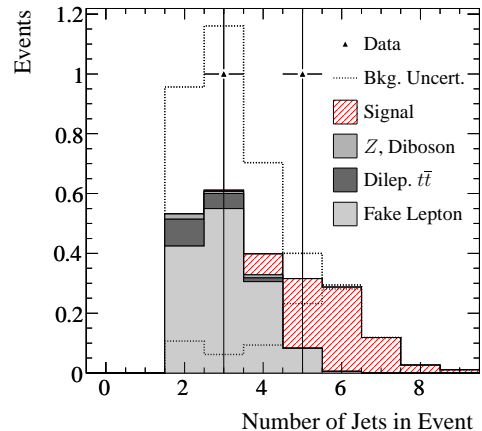}
\caption{Number of reconstructed jets for the expected backgrounds.
  The observed data and the \bprime\ (or $B$) signal are shown at the
  best-fit rate for $m_Q=330$ GeV/$c^2$. The fitted size and shape for
the $\TFT+B$ signal is nearly identical.
\vspace{-3ex}}
\label{fig:data_finalcut_bp}
\end{figure}

\begin{figure}[t]
\includegraphics[width=\linewidth]{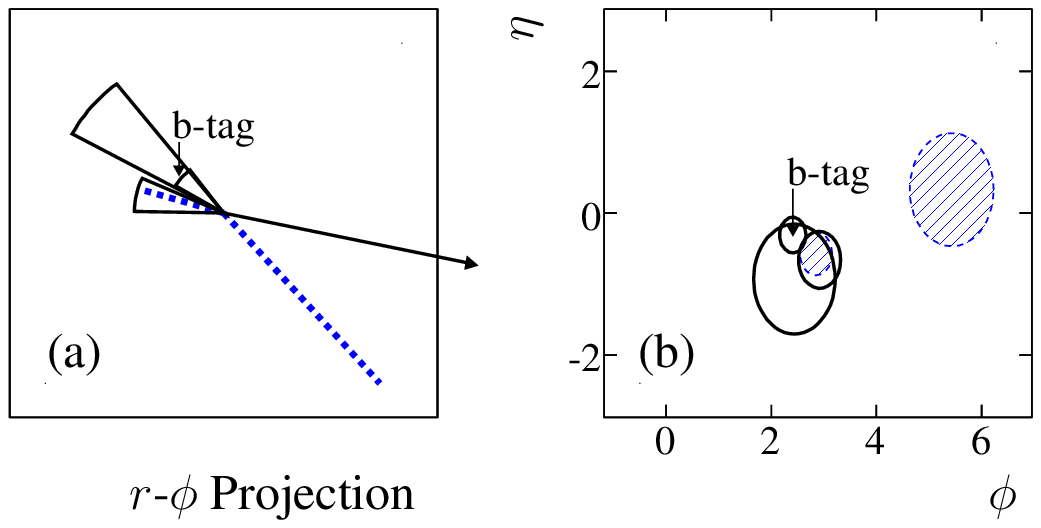}\\
\includegraphics[width=\linewidth]{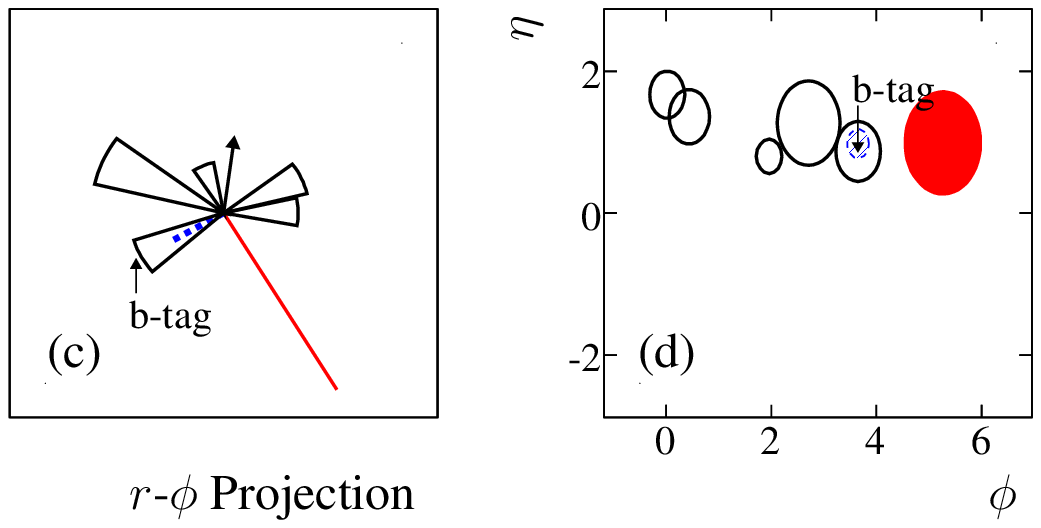}\\
\caption{Event displays for the observed three-jet, $\mu\mu$ event (a,b)
and the five-jet, $e\mu$ event (c,d).
Shown in (a) and (c) are views of the events along the beam axis; jets
  shown as cones, electrons as solid lines, muons as dotted lines
  and missing transverse energy as an arrow; lengths are proportional to
  $p_T$ (see Table~\ref{table:evdetails}). Shown in (b) and (d) are views of the
  events in $\eta-\phi$;  jets shown as open circles, electrons as
  filled circles and muons as dashed circles; radii are proportional
  to $p_T$.
\vspace{-3ex}}
\label{fig:events}
\end{figure}

The $t\overline{t}\rightarrow \ell^+\nu b\ell^-\nu \bar{b}$ backgrounds are
estimated using events generated in {\sc pythia} 6.216 at $m_t = $ 172.5\gevcc,
assuming a $t\overline{t}$ production cross-section of 7.2\,pb.   Modeling of the $t\overline{t}$ contribution is validated
by comparing predicted and observed rates of events with opposite-charge leptons,
large $\missET$, and at least one $b$-tagged jet, where  $t\overline{t}$  is expected to
dominate.

Backgrounds to the $\ell^{\pm}\ell^{\pm}bj\missET$ signature with real
same-charge leptons are rare in the SM; they are largely from $W\!Z$ and
$Z\!Z$ production and are highly suppressed by the requirement of a
$b$-tag.  Backgrounds from diboson production $W\!W$, $ W\!Z$, $Z\!Z$, $W\gamma$, and
$Z\gamma$ in association with $b$ jets are modeled with {\sc pythia} 6.216
and {\sc baur}~\cite{baur} generators.  

Backgrounds from charge mismeasurement are insignificant, as the
charge of a particle with $p_T \approx$ 100\gevc\ is typically
determined with more than 5$\sigma$ significance~\cite{wandz}. Charge
mismeasurement is very rare in this range, confirmed by the
absence of any strong features in dilepton invariant mass in the
$Z$ mass region in same-charge muon events. The largest potential source comes from $t\bar{t}$
events, in which the lepton momenta are typically smaller than
100\gevc. The final background estimates are given in Table~\ref{table:back}.

The \bprime\ and $\TFT\ +B$ signals are modeled with the {\sc{madgraph}}
simulation program following the minimal composite Higgs model described in \cite{Contino:2008hi}
and paired with {\sc pythia} for hadronization and fragmentation.  The acceptance
is approximately 2.2\%, nearly independent of heavy quark masses in the range 300--400 \gevcc.
The expected numbers of events for \bprime (or $B$), and
\TFT\ $+B$ are given in Table~\ref{table:bprres}.

\begin{table}[b]
\vspace{-3ex}
\caption{Transverse momentum (in \gevcnospace) of leptons and
  transverse energy (in \gev) of jets in the two events with the $\ell^{\pm}\ell^{\pm}bj\missET$
signature.}
\begin{ruledtabular}
\begin{tabular}{lrrrrrr}
Event & $\ell_1$ & $\ell_2$ & jet$_1$ & $b$-jet & $\missET$ & other jets  \\ \hline
$\mu^+\mu^+$ & 80 & 31 & 78 & 25 & 87 & 40 \\
$e^+\mu^+$ & 73 & 21 & 60 & 42 & 27 & 39, 33, 24 \\
\end{tabular}
\label{table:evdetails}
\end{ruledtabular}
\end{table}

We observe two events in the signal region, in agreement with the expected backgrounds (see
Table~\ref{table:back}). To calculate the most likely signal cross section,
we perform a binned maximum-likelihood fit
to the number of reconstructed jets.
Figure~\ref{fig:data_finalcut_bp} shows the number of reconstructed
jets in the observed events, as well as the signal distribution with
the best-fit value of the signal cross section.  Kinematics of the two
signal events is shown in Fig.~\ref{fig:events} and the $p_T$ values
are given in Table~\ref{table:evdetails}.

\begin{figure}[t]
\includegraphics[width=0.9\linewidth]{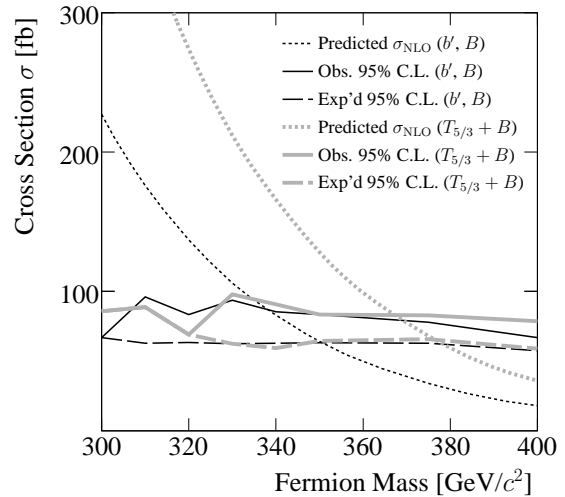}
\caption{Theoretical cross sections for \bprime\ (or $B$)
and $\TFT+B$ with
expected and observed 95\% C.L. limits overlaid.
\vspace{-3ex}}
\label{fig:sensitivity}
\end{figure}

We construct confidence intervals~\cite{feldcous} in the theoretical
cross section
by generating ensembles of simulated experiments that describe
expected fluctuations of statistical and systematic uncertainties,
including uncertainties in the jet-energy scale~\cite{jes},
gluon radiation~\cite{rad}, signal and background normalization,
and parton distribution functions~\cite{pdfs1,pdfs2}.  The median expected and observed limits along with the theoretical
next-to-leading-order (NLO) cross section~\cite{nloa,nlob} are given in Table~\ref{table:bprres} and shown in Fig.~\ref{fig:sensitivity}.

We convert limits on the pair-production cross sections to limits on the
fermion masses and obtain $m_{b^{\prime}}, m_{B} >$ 338\,$\mathrm{GeV}/c^2$ and
$m_{T_{5/3}} >$ 365\,$\mathrm{GeV}/c^2$ at 95\% confidence level.
The two events observed are consistent with the predicted number of background
events, although we note that the $e\mu$ event has a number of jets
characteristic of the signal, reducing the observed lower limits from what is expected.
This is the most restrictive direct lower limit on the mass of a
down-type fourth-generation quark, significantly reducing
the allowed SM mass range, and the first
lower limits on the masses of an exotic doublet $\TFT+B$, which may figure
prominently in future searches.

\input acknow_PRL


\end{document}

%% file: October2009_Authors.tex
\affiliation{Institute of Physics, Academia Sinica, Taipei, Taiwan 11529, Republic of China} 
\affiliation{Argonne National Laboratory, Argonne, Illinois 60439} 
\affiliation{University of Athens, 157 71 Athens, Greece} 
\affiliation{Institut de Fisica d'Altes Energies, Universitat Autonoma de Barcelona, E-08193, Bellaterra (Barcelona), Spain} 
\affiliation{Baylor University, Waco, Texas  76798} 
\affiliation{Istituto Nazionale di Fisica Nucleare Bologna, $^{cc}$University of Bologna, I-40127 Bologna, Italy} 
\affiliation{Brandeis University, Waltham, Massachusetts 02254} 
\affiliation{University of California, Davis, Davis, California  95616} 
\affiliation{University of California, Los Angeles, Los Angeles, California  90024} 
\affiliation{University of California, San Diego, La Jolla, California  92093} 
\affiliation{University of California, Santa Barbara, Santa Barbara,California 93106} 
\affiliation{University of California, Irvine, Irvine, California 92697}
\affiliation{Instituto de Fisica de Cantabria, CSIC-University of Cantabria, 39005 Santander, Spain} 
\affiliation{Carnegie Mellon University, Pittsburgh, PA  15213} 
\affiliation{Enrico Fermi Institute, University of Chicago, Chicago,Illinois 60637}
\affiliation{European Organization for Nuclear Research, Geneva 23, Switzerland}
\affiliation{Comenius University, 842 48 Bratislava, Slovakia; Institute of Experimental Physics, 040 01 Kosice, Slovakia} 
\affiliation{Joint Institute for Nuclear Research, RU-141980 Dubna, Russia} 
\affiliation{Duke University, Durham, North Carolina  27708} 
\affiliation{Fermi National Accelerator Laboratory, Batavia, Illinois 60510} 
\affiliation{University of Florida, Gainesville, Florida  32611} 
\affiliation{Laboratori Nazionali di Frascati, Istituto Nazionale di Fisica Nucleare, I-00044 Frascati, Italy} 
\affiliation{University of Geneva, CH-1211 Geneva 4, Switzerland} 
\affiliation{Glasgow University, Glasgow G12 8QQ, United Kingdom} 
\affiliation{Harvard University, Cambridge, Massachusetts 02138} 
\affiliation{Division of High Energy Physics, Department of Physics, University of Helsinki and Helsinki Institute of Physics, FIN-00014, Helsinki, Finland} 
\affiliation{University of Illinois, Urbana, Illinois 61801} 
\affiliation{The Johns Hopkins University, Baltimore, Maryland 21218} 
\affiliation{Institut f\"{u}r Experimentelle Kernphysik, Karlsruhe Institute of Technology, D-76131 Karlsruhe, Germany} 
\affiliation{Center for High Energy Physics: Kyungpook National University, Daegu 702-701, Korea; Seoul National University, Seoul 151-742, Korea; Sungkyunkwan University, Suwon 440-746, Korea; Korea Institute of Science and Technology Information, Daejeon 305-806, Korea; Chonnam National University, Gwangju 500-757, Korea; Chonbuk National University, Jeonju 561-756, Korea} 
\affiliation{Ernest Orlando Lawrence Berkeley National Laboratory, Berkeley, California 94720} 
\affiliation{University of Liverpool, Liverpool L69 7ZE, United Kingdom} 
\affiliation{University College London, London WC1E 6BT, United Kingdom} 
\affiliation{Centro de Investigaciones Energeticas Medioambientales y Tecnologicas, E-28040 Madrid, Spain} 
\affiliation{Massachusetts Institute of Technology, Cambridge, Massachusetts  02139} 
\affiliation{Institute of Particle Physics: McGill University, Montr\'{e}al, Qu\'{e}bec, Canada H3A~2T8; Simon Fraser University, Burnaby, British Columbia, Canada V5A~1S6; University of Toronto, Toronto, Ontario, Canada M5S~1A7; and TRIUMF, Vancouver, British Columbia, Canada V6T~2A3} 
\affiliation{University of Michigan, Ann Arbor, Michigan 48109} 
\affiliation{Michigan State University, East Lansing, Michigan  48824}
\affiliation{Institution for Theoretical and Experimental Physics, ITEP, Moscow 117259, Russia} 
\affiliation{University of New Mexico, Albuquerque, New Mexico 87131} 
\affiliation{Northwestern University, Evanston, Illinois  60208} 
\affiliation{The Ohio State University, Columbus, Ohio  43210} 
\affiliation{Okayama University, Okayama 700-8530, Japan} 
\affiliation{Osaka City University, Osaka 588, Japan} 
\affiliation{University of Oxford, Oxford OX1 3RH, United Kingdom} 
\affiliation{Istituto Nazionale di Fisica Nucleare, Sezione di Padova-Trento, $^{dd}$University of Padova, I-35131 Padova, Italy} 
\affiliation{LPNHE, Universite Pierre et Marie Curie/IN2P3-CNRS, UMR7585, Paris, F-75252 France} 
\affiliation{University of Pennsylvania, Philadelphia, Pennsylvania 19104}
\affiliation{Istituto Nazionale di Fisica Nucleare Pisa, $^{ee}$University of Pisa, $^{ff}$University of Siena and $^{gg}$Scuola Normale Superiore, I-56127 Pisa, Italy} 
\affiliation{University of Pittsburgh, Pittsburgh, Pennsylvania 15260} 
\affiliation{Purdue University, West Lafayette, Indiana 47907} 
\affiliation{University of Rochester, Rochester, New York 14627} 
\affiliation{The Rockefeller University, New York, New York 10021} 
\affiliation{Istituto Nazionale di Fisica Nucleare, Sezione di Roma 1, $^{hh}$Sapienza Universit\`{a} di Roma, I-00185 Roma, Italy} 

\affiliation{Rutgers University, Piscataway, New Jersey 08855} 
\affiliation{SLAC National Accelerator Laboratory, Menlo Park, California 94025}
\affiliation{Texas A\&M University, College Station, Texas 77843} 
\affiliation{Istituto Nazionale di Fisica Nucleare Trieste/Udine, I-34100 Trieste, $^{ii}$University of Trieste/Udine, I-33100 Udine, Italy} 
\affiliation{University of Tsukuba, Tsukuba, Ibaraki 305, Japan} 
\affiliation{Tufts University, Medford, Massachusetts 02155} 
\affiliation{Waseda University, Tokyo 169, Japan} 
\affiliation{Wayne State University, Detroit, Michigan  48201} 
\affiliation{University of Wisconsin, Madison, Wisconsin 53706} 
\affiliation{Yale University, New Haven, Connecticut 06520} 
\author{T.~Aaltonen}
\affiliation{Division of High Energy Physics, Department of Physics, University of Helsinki and Helsinki Institute of Physics, FIN-00014, Helsinki, Finland}
\author{J.~Adelman}
\affiliation{Enrico Fermi Institute, University of Chicago, Chicago, Illinois 60637}
\author{B.~\'{A}lvarez~Gonz\'{a}lez$^v$}
\affiliation{Instituto de Fisica de Cantabria, CSIC-University of Cantabria, 39005 Santander, Spain}
\author{S.~Amerio$^{dd}$}
\affiliation{Istituto Nazionale di Fisica Nucleare, Sezione di Padova-Trento, $^{dd}$University of Padova, I-35131 Padova, Italy} 

\author{D.~Amidei}
\affiliation{University of Michigan, Ann Arbor, Michigan 48109}
\author{A.~Anastassov}
\affiliation{Northwestern University, Evanston, Illinois  60208}
\author{A.~Annovi}
\affiliation{Laboratori Nazionali di Frascati, Istituto Nazionale di Fisica Nucleare, I-00044 Frascati, Italy}
\author{J.~Antos}
\affiliation{Comenius University, 842 48 Bratislava, Slovakia; Institute of Experimental Physics, 040 01 Kosice, Slovakia}
\author{G.~Apollinari}
\affiliation{Fermi National Accelerator Laboratory, Batavia, Illinois 60510}
\author{A.~Apresyan}
\affiliation{Purdue University, West Lafayette, Indiana 47907}
% ...

\collaboration{CDF Collaboration\footnote{With visitors from $^a$University of Massachusetts Amherst, Amherst, Massachusetts 01003,
$^b$Universiteit Antwerpen, B-2610 Antwerp, Belgium, 
$^c$University of Bristol, Bristol BS8 1TL, United Kingdom,
$^d$Chinese Academy of Sciences, Beijing 100864, China, 
$^e$Istituto Nazionale di Fisica Nucleare, Sezione di Cagliari, 09042 Monserrato (Cagliari), Italy,
$^f$University of California Irvine, Irvine, CA  92697, 
$^g$University of California Santa Cruz, Santa Cruz, CA  95064, 
$^h$Cornell University, Ithaca, NY  14853, 
$^i$University of Cyprus, Nicosia CY-1678, Cyprus, 
$^j$University College Dublin, Dublin 4, Ireland,
$^k$University of Edinburgh, Edinburgh EH9 3JZ, United Kingdom, 
$^l$University of Fukui, Fukui City, Fukui Prefecture, Japan 910-0017
$^m$Kinki University, Higashi-Osaka City, Japan 577-8502
$^n$Universidad Iberoamericana, Mexico D.F., Mexico,
$^o$University of Iowa, Iowa City, IA  52242,
$^p$Kansas State University, Manhattan, KS 66506
$^q$Queen Mary, University of London, London, E1 4NS, England,
$^r$University of Manchester, Manchester M13 9PL, England,
$^s$Muons, Inc., Batavia, IL 60510, 
$^t$Nagasaki Institute of Applied Science, Nagasaki, Japan, 
$^u$University of Notre Dame, Notre Dame, IN 46556,
$^v$University de Oviedo, E-33007 Oviedo, Spain, 
$^w$Texas Tech University, Lubbock, TX  79609, 
$^x$IFIC(CSIC-Universitat de Valencia), 56071 Valencia, Spain,
$^y$Universidad Tecnica Federico Santa Maria, 110v Valparaiso, Chile,
$^z$University of Virginia, Charlottesville, VA  22906
$^{aa}$Bergische Universit\"at Wuppertal, 42097 Wuppertal, Germany,
$^{bb}$Yarmouk University, Irbid 211-63, Jordan
$^{jj}$On leave from J.~Stefan Institute, Ljubljana, Slovenia, 
}}
\noaffiliation

%% file: abstract.tex
We report the most restrictive direct limits on masses of
fourth-generation down-type quarks $b^{\prime}$, and quark-like composite fermions ($B$ or $T_{5/3}$), decaying promptly to
$t W^{\mp}$. We search
for a significant excess of events with two same-charge leptons
($e$, $\mu$), several hadronic jets, and missing transverse energy.
An analysis of data from $p\overline{p}$ collisions with an integrated
luminosity of 2.7\,fb$^{-1}$ collected with the CDF II detector at
Fermilab yields no evidence for such a signal, setting mass limits $m_{b^{\prime}}, m_{B} >$ 338\,$\mathrm{GeV}/c^2$ and
$m_{T_{5/3}} >$ 365\,$\mathrm{GeV}/c^2$ at 95\% confidence level.

%% file: acknow_PRL.tex
We thank the Fermilab staff and the technical staffs of the participating institutions for their vital contributions. This work was supported by the U.S. Department of Energy and National Science Foundation; the Italian Istituto Nazionale di Fisica Nucleare; the Ministry of Education, Culture, Sports, Science and Technology of Japan; the Natural Sciences and Engineering Research Council of Canada; the National Science Council of the Republic of China; the Swiss National Science Foundation; the A.P. Sloan Foundation; the Bundesministerium f\"ur Bildung und Forschung, Germany; the World Class University Program, the National Research Foundation of Korea; the Science and Technology Facilities Council and the Royal Society, UK; the Institut National de Physique Nucleaire et Physique des Particules/CNRS; the Russian Foundation for Basic Research; the Ministerio de Ciencia e Innovaci\'{o}n, and Programa Consolider-Ingenio 2010, Spain; the Slovak R\&D Agency; and the Academy of Finland.

%% file: paper.bbl
\begin{thebibliography}{99}
\bibitem{coordinates}
CDF uses a cylindrical coordinate system with
  the $z$ axis along the proton beam axis. Pseudorapidity is $\eta
  \equiv -\ln(\tan (\theta/2))$, where $\theta$ is the polar angle
  relative to the proton beam direction, and $\phi$ is the azimuthal angle  while $p_T = |p|\ \sin\theta$, $E_T = E\ \sin\theta$.
\bibitem{isodef} A lepton is isolated if the calorimeter energy in a cone $\Delta
  R < 0.4$ surrounding the lepton is less than 10\% of the energy of
  the lepton.
\bibitem{metdef}
Missing transverse energy, $\missET$, is defined as the magnitude of the
vector $ -\sum_i E_T^i \vec n_i$ where $E_T^i$ are the magnitudes of
transverse energy contained in each calorimeter tower $i$, and $\vec n_i$
is the unit vector from the interaction vertex to the tower in the
transverse ($x,y$) plane. $\missET$ is further corrected for the
energy of identified muons.
\end{thebibliography}
